# Cu-doping effects on the ferromagnetic semimetal CeAuGe

Soohyeon Shin[1,*], Vladimir Pomjakushin[2], Marek Bartkowiak[3], Marisa Medarde[1], Tian Shang[4,1], Dariusz J. Gawryluk[1], and Ekaterina Pomjakushina[1,†]

[1] Laboratory for Multiscale Materials Experiments, Paul Scherrer Institut, CH-5232 Villigen PSI, Switzerland

[2] Laboratory for Neutron Scattering and Imaging, Paul Scherrer Institut, CH-5232, Villigen PSI, Switzerland

[3] Laboratory for Neutron and Muon Instrumentation, Paul Scherrer Institut, CH-5232 Villigen PSI, Switzerland

[4] Key Laboratory of Polar Materials and Devices (MOE), School of Physics and Electronic Science, East China Normal University, Shanghai 200241, China

Corresponding authors: [*]soohyeon.shin@psi.ch, [†]ekaterina.pomjakushina@psi.ch

**Abstract**

We present a study of Cu-substitution effects in 4$f$-Ce intermetallic compound CeAu$_{1-x}$Cu$_x$Ge, with potentially unusual electronic states, in the whole concentration range ($x$ = 0.0 – 1.0). The parent CeAuGe compound, crystallizing in a non-centrosymmetric hexagonal structure, is a ferromagnetic semimetal with Curie temperature 10 K. Cu-doping on Au-site of CeAuGe, CeAu$_{1-x}$Cu$_x$Ge, changes the crystal structure from the non-centrosymmetric ($P6_3mc$) to centrosymmetric ($P6_3/mmc$) space group at the concentration $x$ ~ 0.5, where the $c$-lattice constant has a maximum value. Magnetic susceptibility and electrical resistivity measurements reveal that all Cu-doped compounds undergo magnetic phase transition near 10 K, with the maximum transition temperature of 12 K for $x = 0.5$. The neutron powder diffraction experiments show the ferromagnetic ordering of $Ce^{3+}$ magnetic moments with a value of about 1.2 $\mu_B$ at 1.8 K, oriented perpendicular to the hexagonal $c$-axis. By using symmetry analysis, we have found the solutions for the magnetic structure in the ferromagnetic Shubnikov space groups $Cmc'2_1'$ and $P2_1'/m'$ for $x < 0.5$ and $x \geq 0.5$, respectively. Electrical resistivity $\rho(T)$ exhibits a metallic temperature behaviour in all compositions. The resistivity $\rho(T)$ has a local minimum in the paramagnetic state due to Kondo effects at high doping $x = 0.8$ and 1.0. At the small Cu-doping level, $x = 0.2$, the resistivity shows a broad feature at the ferromagnetic transition temperature and an additional transition-like peculiarity at 2.5 K in the ferromagnetic state.

**1. Introduction**

Ce-based intermetallic compounds, in which localized moments of $Ce^{3+}$ ion form Kondo lattice, exhibit exotic quantum phenomena such as heavy-fermion superconductivity [1-3], topological magnetism [4], intriguing magnetic ordering [5,6], and quantum phase transition [7,8]. The ground states of the Ce-based metals are controlled by competing interactions between the spin anisotropy, multiple exchange interactions, and Kondo hybridization. The Kondo hybridization between localized Ce moment and itinerant spins can be tuned by non-thermal parameters such as pressure, magnetic field, and chemical doping. Chemical doping modifies the density of states at the Fermi level, whereby exchange interactions between localized and itinerant spin change the ground state of the system as described in the Doniach phase diagram [9].

Table 1 summarizes Ce-based hexagonal compounds of Ce$MX$ ($M$ = Cu, Ag, Au, and $X$=Si, Ge, Sn). All systems exhibit magnetically ordered states. Figure 1 shows a linear relation between transition temperatures of magnetic ordering ($T_C$, $T_N$) and an anisotropic parameter $\alpha$ = $d_c/d_a$, when $d_a$ and $d_c$ are Ce-Ce distance along the crystallographic $a$- and $c$-axis, respectively. When the magnetic exchange interactions between Ce-spins are anisotropic, $0.80 < \alpha < 0.875$, the magnetic ground state is antiferromagnetic (AFM) one, while isotropic interactions, $0.875 < \alpha < 0.95$, favour a ferromagnetic (FM) ordering. As shown in Table 1, all Ce$MX$ compounds have similar crystalline-electric-field ground states (GS$_{CEF}$) of mainly $|\pm1/2\rangle$ with mixing of $|\mp5/2\rangle$ doublet.

Table 1. Summary of crystal structure and magnetic properties of Ce$MX$ ($M$ = Cu, Ag, Au, and $X$ = Si, Ge, Sn) intermetallic compounds

|  | $M$ | Space group | Type | $T_C$, $T_N$ (K) | GS$_{CEF}$ | $a$ (Å) | $c$ (Å) | $d_c/d_a$ | Ref. |
|---|---|---|---|---|---|---|---|---|---|
| Ce$M$Si | Cu | $P6_3/mmc$ | FM | 15.5 | $0.87p+0.48q$ | 4.238 | 7.988 | 0.940 | [10,11] |
| Ce$M$Ge | Cu | $P6_3/mmc$ | FM | 10.2 | $0.99p+0.01q$ | 4.311 | 7.933 | 0.920 | [12,13] |
|  | Ag | $P6_3mc$ | AFM | 4.8 | - | 4.544 | 7.711 | 0.850 | [14,15] |
|  |  | $P6_3/mmc$ | FM | 6.0 |  | 4.536 | 7.746 |  |  |
|  | Au | $P6_3mc$ | FM | 9.7 | - | 4.460 | 7.936 | 0.890 | [16] |
| Ce$M$Sn | Cu | $P6_3mc$ | AFM | 8.6 | $0.96p+0.27q$ [a] | 4.583 | 7.865 | 0.860 | [12,17] |
|  | Ag | $P6_3mc$ | AFM | 6.45 | $0.93p+0.35q$ [a] | 4.77 | 7.74 | 0.810 | [17,18] |
|  | Au | $P6_3mc$ | AFM | 4.1 | $0.93p+0.35q$ [a] | 4.72 | 7.70 | 0.815 | [17,18] |

$T_C$, $T_N$: magnetic ordering temperature, $d_c$ ($d_a$): distance between Ce atoms along the $c$-axis ($a$-axis) direction. GS$_{CEF}$: crystalline-electric-field ground state with $p = |1/2\rangle$ and $q = |5/2\rangle$. [a] Mantid software [19] was used for simulating GS$_{CEF}$ with the $B_n^m$ crystal field parameters given in the references.

An unusual non-collinear FM structure has been reported in CeAlSi, presumably due to the non-centrosymmetric crystal structure [20]. In our case, both non-centrosymmetric end member CeAuGe and the centrosymmetric one CeCuGe exhibit ferromagnetism with transition temperature near 10 K. This is in accordance with a previous neutron powder diffraction study that has shown that the magnetic ordering in CeAuGe is a collinear FM, and we note that the magnetic structure of CeCuGe has not been reported [21]. Cu-doping of the non-centrosymmetric CeAuGe is the way of investigating the relationship between the magnetic ordering, crystal symmetry, and the bulk physical properties, while the Curie temperature remains similar. Herein, we report the crystal and magnetic structures of the polycrystalline series of CeAu$_{1-x}$Cu$_x$Ge using x-ray and neutron diffraction and the macroscopic physical properties such as electrical resistivity and magnetization as a function of temperature and magnetic field.

## 2. Material and methods

Polycrystalline samples of CeAu$_{1-x}$Cu$_x$Ge were synthesized by the arc-melting technique, and then the arc-melted samples were sealed in the evacuated silica tubes with a Ta-foil then the tubes were quenched in the cold water after annealed at 800 °C for 10 days to improve the homogeneity of the samples. Phase purity and crystal structures of CeAu$_{1-x}$Cu$_x$Ge were checked by powder x-ray diffraction measurements (PXRD) using a Bruker D8 Advance with Cu-cathode. Chemical composition was confirmed by x-ray micro-fluorescence (XRF) analysis using an Orbis micro-XRF analyzer from EDAX. The crystal and magnetic structures were studied by neutron powder diffraction using the high-resolution powder diffractometer for thermal neutrons (HRPT) [22] at the Swiss Spallation Neutron Source SINQ at Paul Scherrer Institut (PSI), Switzerland. About 2.5 g of each powder was loaded in 6 mm vanadium containers. Diffraction patterns were collected at temperatures of 1.5 and 15 K using neutrons with wavelengths of 1.494 and 2.45 Å. All diffraction data were analyzed using the programs of the FullProf software suite [23]. The symmetry analysis of the magnetic structures was done using the Bilbao crystallographic server [24] and the ISODISTORT tool based on ISOTROPY software [25,26]. Electrical resistivity measurements were performed using the standard four-probe (25 μm Pt wires) technique applying a current of 1 mA on the polished surface of bar-shaped specimens. A physical property measurement system (PPMS-9, Quantum Design) was used for applying magnetic fields up to 90 kOe and controlling the temperature in the range from 1.8 to 300 K. A $^3$He-pumping system with an Oxford vertical magnet cryostat was adopted for the resistivity measurements in the temperature range from 0.28 to 2 K. The magnetization measurements were performed on a superconducting quantum interference device (SQUID) installed in the magnetic property measurement system (MPMS-7, Quantum Design), in the temperature and magnetic field ranges from 1.8 to 300 K and 0 to 70 kOe, respectively.

## 3. Results

### 3.1. Crystal structure of CeAu$_{1-x}$Cu$_x$Ge

Figure 2(a) shows crystal structures of CeAuGe and CeCuGe and representative PXRD patterns for CeAu$_{1-x}$Cu$_x$Ge ($x$ = 0.0, 0.5, 1.0). Both end-member compounds adopt hexagonal structure, wherein the two different hexagonal layers of Ce – Ce and $M$ – Ge ($M$ = Cu and Au) stack alternately along the crystallographic $c$-axis. While the puckered hexagonal layer of Au – Ge breaks the centrosymmetric symmetry ($P6_3mc$, no. 186), the flat Cu – Ge layer conserves the centrosymmetric symmetry ($P6_3/mmc$, no. 194) of the hexagonal structure [12,27]. Figure 2(b) and (c) show lattice parameters of CeAu$_{1-x}$Cu$_x$Ge determined from so-called Le Bail fits of the x-ray diffraction patterns using the $P6_3mc$ and $P6_3/mmc$ space groups for $0 \leq x \leq 0.5$ and $0.5 < x < 1.0$ compounds, respectively. In the Le Bail fit, only the crystal metrics and resolution parameters are refined, whereas the integrated Bragg peak intensities are refined independently without any structure model. The crystal structures were determined from the neutron diffraction data (see below). Cu-doping ($x$) linearly shortens the $a$-lattice parameter as $x$ increases, decreasing about 3 % at $x$ = 1. On the other hand, the $c$-lattice parameter exhibits two different linear regimes with an abrupt change near $x = 0.5$, which can be interpreted by the different crystal symmetry. The unit cell volume linearly decreases as $x$ increases because the $a$-lattice parameter change is dominant.

### 3.2. Magnetic properties of CeAu$_{1-x}$Cu$_x$Ge

Magnetic susceptibility ($M/H$) of CeAu$_{1-x}$Cu$_x$Ge is plotted as a function of temperature in Fig. 3(a), and all compounds show FM-like phase transition. Figure 3(b) shows that the magnetic susceptibility in all samples well follows the Curie-Weiss behaviour at temperatures above 100 K. From the least-squares fits using an equation of $M/H = M/H(0) + C/(T - \Theta)$ ($M/H(0)$: temperature-independent term, $C$: Curie-constant, $\Theta$: Curie-Weiss temperature), the deduced effective magnetic moment ($\mu_{eff}$) is plotted as a function of $x$ in the inset of Fig. 3(b). All $\mu_{eff}$ values are close to 2.54 $\mu_B$, which is the expected value for $J=5/2$ of Ce$^{3+}$. The deduced Curie-Weiss temperatures ($\Theta$) are of negative sign, except for $x = 1.0$, at temperatures above 100 K. The negative $\Theta$ at high temperatures is due to the anisotropic magnetic exchange interaction whereby the $\Theta$ deduced by the field applied along the $c$-axis is -78(1) K ($\Theta = 2.7(5)$ K for $ab$-plane) from the single-crystal experiments [28]. As shown in Table 2, the deduced $\Theta$ from the fit in temperature range from 90 to 30 K is of positive sign and comparable to the Curie temperature. Figure 3(c) shows the magnetic field dependence of magnetization $M(H)$ of CeAu$_{1-x}$Cu$_x$Ge at temperatures 2 and 20 K. The magnetic moments of CeAu$_{1-x}$Cu$_x$Ge at 70 kOe are about ~1 $\mu_B$ at 2 K and it is consistent with the calculated value of 1.05 $\mu_B$ [29] using the CeCuGe CEF ground state of 0.99|±1/2> + 0.01|∓5/2> [13], implying that all samples in the series might have the similar CEF ground state, mainly |±1/2> doublet. All samples show paramagnetic behaviour with small FM fluctuations of $M(H)$ at 20 K. The temperature, where the derivative of magnetization $\partial M/\partial T$ shows a minimum as depicted in the inset of Fig. 3(d), was assigned to the critical temperature of the ferromagnetic transition ($T_C$). Figure 3(d) shows the maximum $T_C$ near $x \sim 0.5$ in the $T - x$ axes where the $c$-lattice parameter has a maximum value.

Table 2. Curie-Weiss fit results of CeAu$_{1-x}$Cu$_x$Ge at two temperature ranges using the equation of $M/H = M/H(0) + C/(T - \Theta)$.

| $x$ | $T = 300 - 100$ K | | | $T = 90 - 30$ K | | |
|---|---|---|---|---|---|---|
| | $M/H(0)$ (emu/mol) | $\mu_{eff}$ ($\mu_B$) | $\Theta$ (K) | $M/H(0)$ (emu/mol) | $\mu_{eff}$ ($\mu_B$) | $\Theta$ (K) |
| 0.0 | -1.48(16) x10$^{-4}$ | 2.68(2) | -22.6(1) | 2.50(2) x10$^{-3}$ | 1.83(1) | 10.8(1) |
| 0.2 | -1.57(17) x10$^{-4}$ | 2.82(2) | -82.9(15) | 2.62(08) x10$^{-3}$ | 1.40(03) | 11.3(2) |
| 0.5 | 4.89(48) x10$^{-6}$ | 2.55(5) | -5.1(21) | 1.83(6) x10$^{-3}$ | 2.06(1) | 11.6(1) |
| 0.8 | 2.33(16) x10$^{-4}$ | 2.47(2) | -0.7(8) | 1.63(5) x10$^{-3}$ | 2.10(1) | 11.0(3) |
| 1.0 | 1.15(44) x10$^{-4}$ | 2.47(6) | 5.5(26) | 1.43(25) x10$^{-3}$ | 2.26(7) | 10.3(5) |

### 3.3. Neutron powder diffraction results of CeAu$_{1-x}$Cu$_x$Ge

Magnetic and crystal structures of CeAu$_{1-x}$Cu$_x$Ge ($x$ = 0.0, 0.2, 0.5, 0.8, 1) were studied using neutron powder diffraction. Table 3 summarizes the results of the Rietveld refinements of the crystal structures from the neutron diffraction patterns measured with neutron wavelength $\lambda$ = 1.494 Å at 15 K. The best fits were obtained in the space groups $P6_3mc$ and $P6_3/mmc$ for $0 \leq x < 0.5$ and $0.5 \leq x < 1.0$, respectively (for details see Fig. S1 in Supplemental Material), in accordance with x-ray diffraction data shown above. For the magnetic structures, the longer wavelength, $\lambda$ = 2.54 Å, was used to obtain higher resolution for the Bragg peaks at low scattering angles $2\theta$. Figure 4(a) and (b) shows representative diffraction patterns of CeAuGe and CeCuGe, respectively. As shown in each upper panel of Fig. 4(a) and (b), the Bragg peak positions at the temperatures 15 and 1.5 K are practically identical, while the Bragg peak intensities at 1.5 K are enhanced. The difference patterns (1.5 – 15 K) for CeAuGe and CeCuGe are shown in the lower panel of Fig 5(a) and (b). Le Bail fits of the difference patterns of CeAu$_{1-x}$Cu$_x$Ge show that the propagation vector $k$ of all compounds is zero, $k$ = 0, with the goodness of fit $\chi^2 \sim 1$ (see details Fig. S2 of SM).

Table 3. Crystal structure parameters refined from the neutron powder diffraction patterns of CeAu$_{1-x}$Cu$_x$Ge measured with neutron wavelength λ = 1.494 Å at 15 K.

| $x$ | Space group | $a$ (Å) | $c$ (Å) | M W, (1/3, 2/3, $z$) | Ge W, (1/3, 2/3, $z$) | $R_{wp}$ | $R_{exp}$ | $\chi^2$ |
|---|---|---|---|---|---|---|---|---|
| 0.0 | $P6_3mc$ | 4.4570(1) | 7.8517(2) | 2b, 0.7640(10) | 2b, 0.2179(11) | 16.9 | 5.40 | 9.73 |
| 0.2 | $P6_3mc$ | 4.4564(2) | 7.8481(4) | 2b, 0.7846(14) | 2b, 0.2415(14) | 30.5 | 8.43 | 13.1 |
| 0.5 | $P6_3/mmc$ | 4.3931(1) | 7.8784(3) | 2c, 0.75 | 2c, 0.25 | 33.0 | 16.4 | 4.04 |
| 0.8 | $P6_3/mmc$ | 4.3420(2) | 7.9252(6) | 2c, 0.75 | 2c, 0.25 | 46.6 | 21.7 | 4.61 |
| 1.0 | $P6_3/mmc$ | 4.2859(3) | 7.9285(6) | 2c, 0.75 | 2c, 0.25 | 27.0 | 16.2 | 2.78 |

$\chi^2 = (R_{wp}/R_{exp})^2$. $\chi^2$, $R_{wp}$, and $R_{exp}$ represent a reduced chi-square, weighted profile factor, and expected weighted profile factor, respectively [23]. The atomic coordinate of Ce 2a (0, 0, $z$) in $P6_3mc$ is fixed to zero for maintaining crystal symmetry. W represents Wyckoff position in relevant crystal symmetry.

### 3.4. Magnetic structure of CeAu$_{1-x}$Cu$_x$Ge

For $P6_3mc$ (0 ≤ $x$ < 0.5) and the propagation vector $k$ = 0, there are four irreducible representations (irreps) for non-zero Ce moments allowed by symmetry. Assuming that the only one irrep is active at the magnetic transition, eight possible magnetic Shubnikov subgroups (labeled as 1, 2, 4–7, 12, and 13 in Fig. S3(a)) are allowed, and four of them (labeled as 1, 2, 5, and 7) have maximal symmetry. For $P6_3/mmc$ (0.5 < $x$ < 1.0) with $k$ = 0, four irreps and eight possible magnetic subgroups (labelled as 1, 2, 4–7, 12, and 13 in Fig. S3(b)) are allowed and four of them (labelled as 1, 2, 5, and 7) have maximal symmetry. All Shubnikov magnetic space groups (MSG) were sorted to fit the difference patterns, starting from maximal symmetry subgroups. The best fit results were obtained in the orthorhombic $Cmc'2_1'$ (no. 36.175, labeled as 5 in Fig. S3(a)) and monoclinic $P2_1'/m'$ (no. 11.54, labeled as 12 in Fig. S3(b)) MSG for CeAuGe and CeCuGe with the goodness of fit $\chi^2$ ~ 1 (see the details in Fig. S4 of SM). The unit cell metrics and atomic coordinates in the fits with the magnetic space groups were fixed by the values refined from the patterns in paramagnetic state at 15K. We note that the goodness of fit $\chi^2$ is the same as for the Le Bail fit, implying that there is no room for further improvement because Le Bail fit is model-independent.

The schematic drawing of the magnetic structures of CeAuGe and CeCuGe is shown in Fig. 4(c) and (d), respectively. The unit cell transformations from parent paramagnetic space groups are given by the following relations **A** = –**b**, **B** = 2**a** + **b**, and **C** = **c** in *Cmc'*2$_1$' and **A** = **a**, **B** = **c**, and **C** = –**b** in *P*2$_1$'/*m*', where the capital letter and lower case are the basis vectors of the magnetic and the parent paramagnetic space group, respectively. Structure parameters obtained from the fits of neutron diffraction patterns are summarised in Tables S2 and S3. In *Cmc'*2$_1$', the Wyckoff positions for Ce is 4a (0, 0, 0) with magnetic moments allowed only along the crystallographic *a*-axis. Thus, the magnetic symmetries impose strong constraints on the possible magnetic configurations. On the other hand, the magnetic moments of Ce in Wyckoff position 2a (0, 0, 0) in *P*2$_1$'/*m*' are allowed along the arbitrary direction. The best result of Rietveld refinement, however, was obtained for the moment directed along the *a*-axis. For both MSGs, the collinear ferromagnetic structure is fixed by symmetry. The refined values of ordered magnetic moments amounted to 1.05(2) $\mu_B$ for CeAuGe and 1.11(3) $\mu_B$ for CeCuGe. They are consistent with the saturated moments in Fig. 3(c) and the previous neutron experiment result of CeAuGe [21]. Figure 4(e) shows the ordered magnetic moments ($\mu_{ord}$) as a function of *x* deduced from the fits of the neutron diffraction patterns (details are given in Table S2 in SM). The samples of CeAu$_{1-x}$Cu$_x$Ge with non-centrosymmetric MSG, $0 \leq x < 0.5$, exhibit the orthorhombic *Cmc'*2$_1$' FM structure, while the compositions with centrosymmetric MSG, $0.5 \leq x \leq 1.0$, adopt *P*2$_1$'/*m*' monoclinic symmetry. We note that the FM structures are similar but have different crystallographic symmetries.

### 3.5. Electrical resistivity of CeAu$_{1-x}$Cu$_x$Ge

Figure 5(a) shows the zero-field electrical resistivity as a function of temperature ($\rho(T)$) of CeAu$_{1-x}$Cu$_x$Ge with showing high resistivity values and metallic behaviour, which are consistent with the previous report [30]. The residual-resistance ratio (RRR) coefficients of

CeAu$_{1-x}$Cu$_x$Ge are of range from 1.2 to 10.6 (details are given in Table S4 in SM). The first-principles calculations show that CeAuGe and CeCuGe both are classified to the semimetal [31]. As shown in the top panel, CeAuGe ($x = 0$) exhibits a kink (red downward arrow) which is consistent with the $T_C$ determined by a dip of $\partial M/\partial T$. Except for $x = 0.2$, $\rho(T)$ in every panel shows the apparent kink at $T_C$. At temperatures above $T_C$, a local minimum ($T_K'$) in $\rho(T)$ is observed in $x = 0.8$ and 1.0 due to the Kondo effects, suggesting that the Kondo hybridization between 4$f^1$ and conduction electrons becomes stronger near $x \sim 1$. The local minimum in $\rho(T)$ is better visually seen in the plot in the overall temperature range, as shown in Fig. S5. Upper panel of Fig. 5(b) shows $\rho(T)$ and $M(H)$ of $x = 0.2$ plotted as a function of temperature on the left and right ordinate, respectively. The transition temperatures of $\rho(T)$ and $M(H)$ were determined by the peak of $\partial M/\partial T$ (black dashed line) and $\partial \rho/\partial T$ (solid red line) for $x = 0.2$ and separated by ~2 K due to the broadened FM phase transition. In addition, the resistivity $\rho(T)$ exhibits a broad hump in the ferromagnetic state around 2.5 K. The lower panel of Fig. 5(b) shows neutron integrated intensity (int. $I$) for $x = 0.2$ as a function of temperature obtained from the Gaussian fits of two most intense magnetic peaks (200) and (001). The FM contribution appears below 8 K, consistent with the magnetization $\partial M/\partial T$. As shown in Fig. 5(c), on the other hand, resistivity exhibits a drop near 12.5 K then moderately decreases from ~8 K, where the bulk $T_C$ is observed. With further lowering the temperature, the transition-like broad hump is observed below 2.5 K, and it decreases down to 0.28 K. When subjected to the magnetic field applied along the out-of-plane direction, the onset temperature of the resistivity drop at 12.5 K is increased and broadened as the field increases, while the transition-like feature around 2.5 K stays robust up to 50 kOe.

## 4. Discussion

Distances between the hexagonal Ce-layers ($d_c$, as defined in Fig. 1) of CeAuGe (3.968(1)

Å) and CeCuGe (3.967(1) Å) are similar. Band structures of ScAuGe, CeAuGe, and LuAuGe calculated using TB-LMTO-ASA (Tight-Binding, Linear Muffin-Tin Orbital, Atomic-Sphere Approximation) program show that the larger spacing $d_c$ leads to the smaller inter-layer interaction [27]. As shown in Fig. 3(d), the maximum $d_c$ is observed at $x\sim0.5$, where the Curie temperature $T_C$ reaches the maximum value of 12 K. The higher $T_C$ = 15.5 K in CeCuSi [10] can be interpreted by the larger $d_c$ = 3.994 Å than the one in CeAuGe and CeCuGe. Similar behaviour was reported in CeCuGe$_{1-x}$Sn$_x$ that the $T_C$ is proportional to the interlayer distance. On the other hand, $T_C$ is inverse proportional to the interlayer distance in CeCu$_{1-x}$Al$_x$Ge, in which Al-doping was expected to screen Ce moments with a higher electron density. $T_C$ in CeAu$_{1-x}$Cu$_x$Ge may be correlated with not only the interlayer interaction but also conduction electron density by the chemical doping.

The similar anisotropic parameters $\alpha = d_c/d_a$ of CeAu$_{1-x}$Cu$_x$Ge are responsible for the similar magnetic properties and collinear ferromagnetic structure. While the magnetic properties of CeAu$_{1-x}$Cu$_x$Ge compounds are similar, the electrical resistivity exhibits different behaviours depending on doping concentration $x$. CeCuGe exhibits the local minimum of $\rho(T)$ at 17.5 K due to the Kondo scattering and $T$-linear behaviour in the FM state. On the other hand, $\rho(T) \sim T^3$ behaviour in the FM state was observed in CeAuGe without the local minimum of $\rho(T)$. Note that the fit by a gap function, $\rho(T) \sim T^n exp(-\Delta/k_B T)$, does not work for both cases. Since the temperature exponent of resistivity in CeCuGe and CeAuGe polycrystals are different from the one expected in the ferromagnetic metals [32,33], $\rho(T)$ measurements on single crystals and at lower temperatures might help to understand the scattering mechanism in the FM state. As shown in Fig. 5(b), the small Cu-doping in CeAuGe suppresses FM transition temperature to 8 K, while other doping cases show $T_C$ higher than $x$ = 0.0 and 1.0. Further experiments with $x$ = 0.2 concentration and other small doping contents can be useful to elucidate the origin of the broad transition at 12.5K and the low-temperature transition-like feature at 2.5 K that are seen only in the electrical resistivity.

5. Conclusion

Cu-doping effects in the hexagonal ferromagnet CeAuGe have been studied on polycrystalline series of CeAu$_{1-x}$Cu$_x$Ge ($x$=0.0, 0.2, 0.4, 0.5, 0.6, 0.8, 1.0) using x-ray and neutron (except $x$ = 0.4 and 0.6) diffraction, electrical resistivity and magnetization as a function of temperature and magnetic field. The doping changes the crystal structure from the non-centrosymmetric ($P6_3mc$) to the centrosymmetric ($P6_3/mmc$) space group at the critical concentration $x_c \sim 0.5$. The magnetic moments of Ce are ferromagnetically ordered in the Shubnikov magnetic space groups $Cmc'2_1'$ and $P2_1'/m'$ below and above $x_c$, respectively. The Curie temperature $T_C$ depends on concentration and reaches the maximum value $T_C$ =12 K at $x$ = $x_c$, where the distance between Ce-layer shows the maximal value. The magnetic moments are oriented perpendicular to the hexagonal $c$-axis with the sizes slightly dependent on concentration amounting to 0.95(2) – 1.16(1) $\mu_B$ at 1.5 K. The electrical resistivity exhibits metallic temperature dependence and a pronounced sharp change at the ferromagnetic transition for all concentrations except for $x$ = 0.2. The resistivity $\rho(T)$ has a local minimum in the paramagnetic state due to Kondo effects at high doping levels $x > 0.5$. The sample with the small doping $x$ = 0.2 shows different from other concentrations behaviour of electrical resistivity. The transition seen by resistivity is broadened and shifted to 12.5 K, which is higher than the bulk $T_C$ and in addition, there is a transition-like feature in the ferromagnetic state at 2.5 K. Further experiments with the low doping content samples can be useful to elucidate the origin of the above peculiarities.

The data that support this study are available via the Zenodo repository [34].

**Acknowledgments**

Work at the Paul Scherrer Institut in Switzerland was supported by SNSF Projects No. 200021_188706 and 206021_139082. Neutron scattering experiments were carried out at the continuous spallation neutron source SINQ at the Paul Scherrer Institut at Villigen PSI in Switzerland. Magnetization measurements were performed at the Laboratory for Multiscale Materials Experiments at the Paul Scherrer Institute. T.S. acknowledges support from the Natural Science Foundation of Shanghai (Grant Nos. 21ZR1420500 and 21JC1402300).

**Figure caption**

Figure 1.

Magnetic ordering temperatures $T_C$, $T_N$ of Ce$MX$ ($M$=Cu, Ag, Au, and $X$=Si, Ge, Sn) as a function of anisotropic parameter $\alpha = d_c/d_a$, where $d_a$ and $d_c$ represent Ce-Ce bond length along the crystallographic *a*- and *c*-axis, respectively. A dashed line is a guide to the eye. The schematic drawing represents a hexagonal crystal structure of Ce$MX$.

Figure 2.

(a) Schematic drawing of the centrosymmetric CeCuGe ($P6_3/mmc$) and non-centrosymmetric CeAuGe ($P6_3mc$) crystal structure (upper panel). Powder X-ray diffraction patterns of CeAu$_{1-x}$Cu$_x$Ge ($x$ = 0, 0.5, 1.0). Black dot, red line, blue and green bar represent observed data, fit result using Le Bail method (explained in the text), nuclear diffraction positions of $P6_3/mmc$ and $P6_3mc$, respectively. (b, c) *a*- and *c*-lattice parameters as a function of $x$ (d) The unit cell volume as a function of $x$. The error bars are smaller than the symbol size.

Figure 3.

(a) Magnetic susceptibility ($M/H$) of CeAu$_{1-x}$Cu$_x$Ge as a function of temperature, the inset shows the data at low temperatures. (b) $H/M$ of CeAu$_{1-x}$Cu$_x$Ge is plotted as a function of temperature, and the inset shows each effective magnetic moment deduced from the Curie-Weiss fit at temperatures above 200 K. (c) Magnetization as a function of field of CeAu$_{1-x}$Cu$_x$Ge at 2 and 20 K. (d) Curie temperature ($T_C$, square symbol) and the room temperature *c*-lattice parameter (circles) as a function of $x$ on the left and right ordinate, respectively. Inset shows the magnetization derivative $\partial M/\partial T$ as a function of temperature, which has a minimum at $T_C$.

Figure 4.

Neutron powder diffraction data collected at 15 and 1.5 K (upper panel) and their difference pattern (lower panel) of CeAuGe (a) and CeCuGe (b) are representatively displayed. The red line in the lower panel represents the best fit results using the Shubnikov magnetic space group $Cmc'2_1'$ for CeAuGe and $P2_1'/m'$ for CeCuGe. The schematic magnetic structures of CeAuGe and CeCuGe are displayed in (c) and (d), respectively. (e) The sizes of ordered magnetic moments obtained from the fits on the difference (1.5 – 15 K) pattern are plotted as a function of $x$.

Figure 5.

(a) Electrical resistivity of $CeAu_{1-x}Cu_xGe$ as a function of temperature, $\rho(T)$, in each panel from top to bottom ($x$ = 0.0, 0.2, 0.5, 0.8, 1.0). Red downward and black upward arrows indicate the Curie temperature and the local minimum in $\rho(T)$. Red solid line represents a least-squares fit using $\rho(T) = \rho_0 + A\,T^n$. (b) $\rho(T)$ and $M/H$ of $x$ = 0.2 are plotted using left and right ordinate, respectively, in the upper panel, and neutron integrated intensity (int.$I$) of the sum of two magnetic peaks (200) and (001) of $x$ = 0.2 is plotted in the lower panel. Dashed black and solid red lines in the upper panel represent the peak of $\partial M/\partial T$ and $\partial \rho/\partial T$, respectively. In the lower panel, the red line shows the best result of least-squares fit from $int.I = bkg + B(1-T_C/T)^{2\nu}$ with bkg = 950, B = 127(10), $T_C$ = 8.5(1), and $\nu$ = 0.26(7). (c) $\rho(T)$ of $x$ = 0.2 was measured under various magnetic fields applied perpendicular to the measured sample surface, as shown in the inset. Black arrows indicate the field dependence of $\rho(T)$ as the field increases.

**Figures**

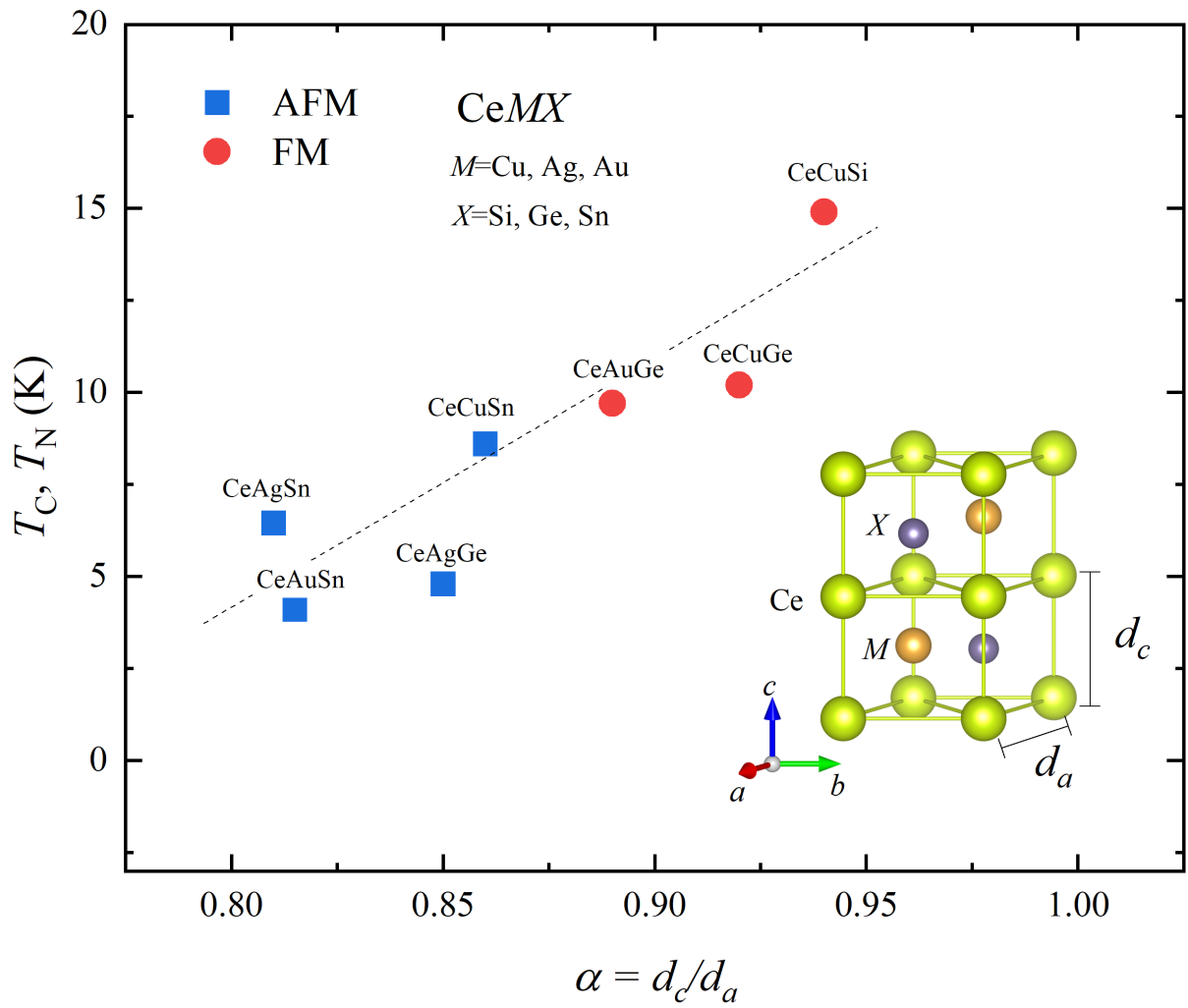

Figure 1

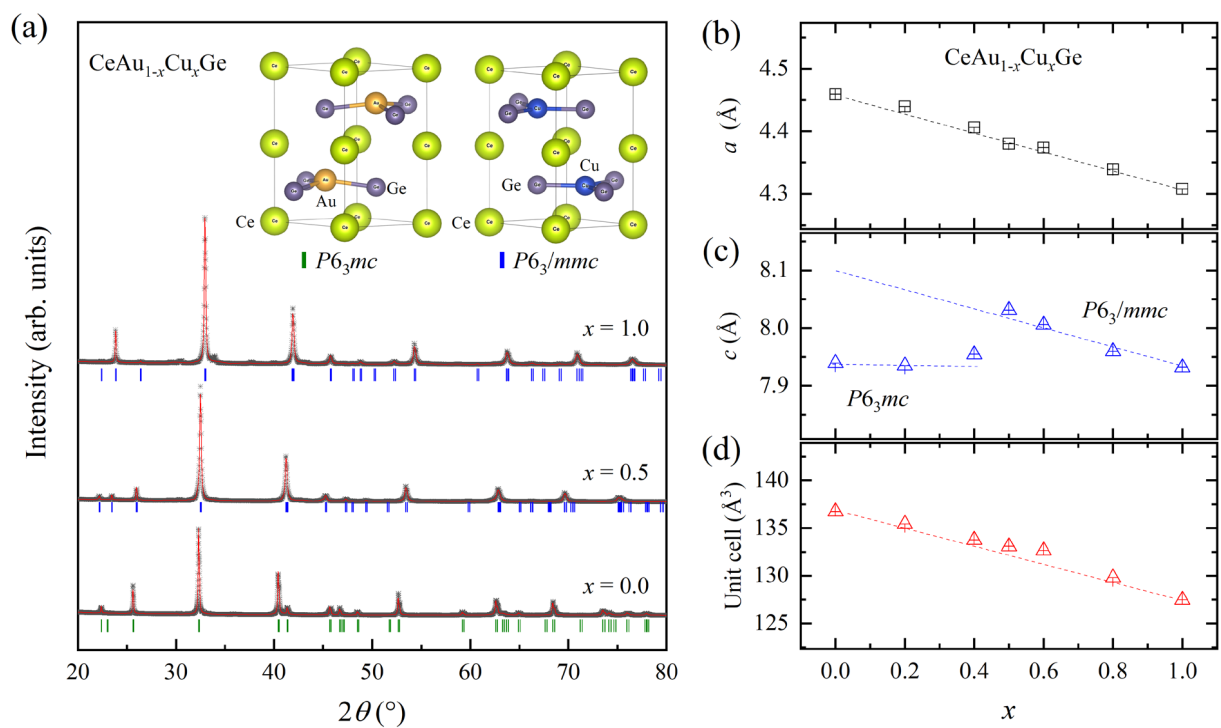

Figure 2

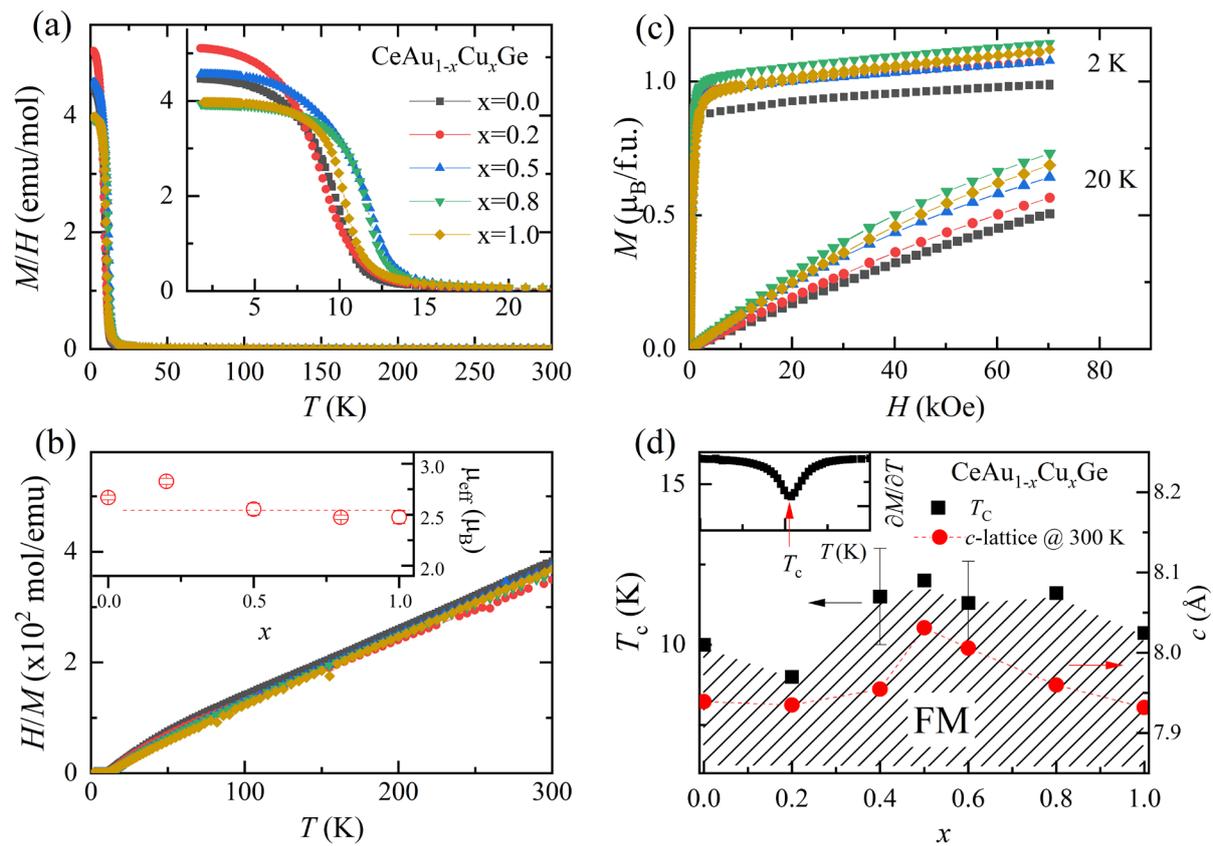

Figure 3

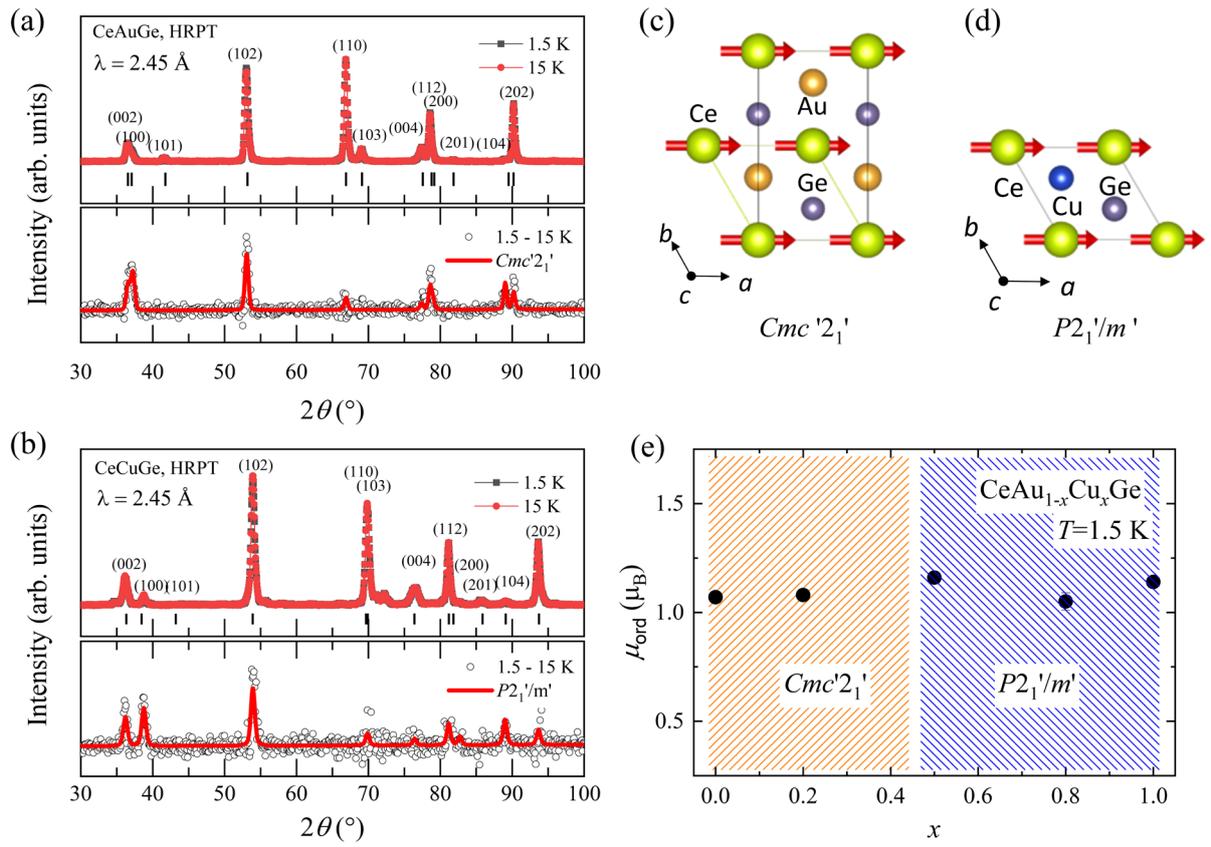

Figure 4

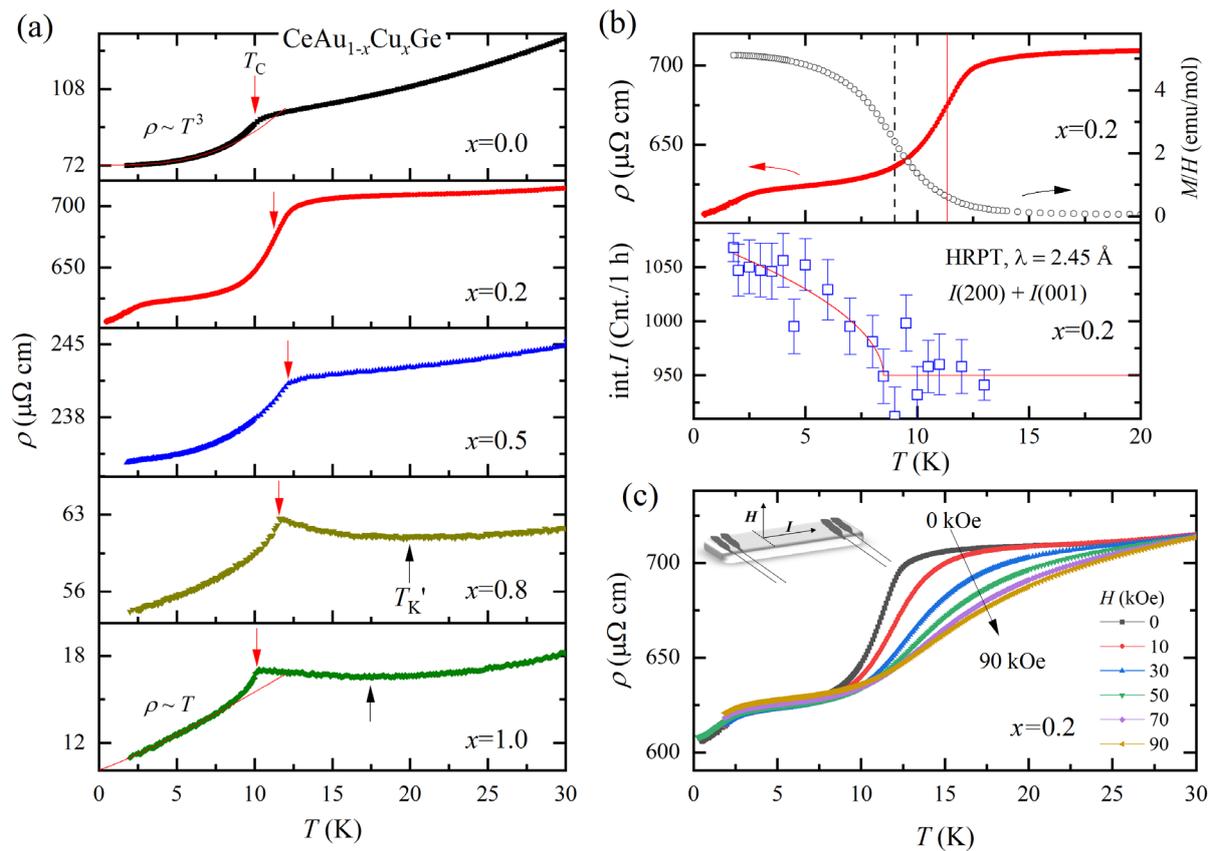

Figure 5

# Supplemental Material for "Cu-doping effects on the ferromagnetic semimetal CeAuGe"

Soohyeon Shin[1,*], Vladimir Pomjakushin[2], Marek Bartkowiak[3], Marisa Medarde[1], Tian Shang[4,1], Dariusz J. Gawryluk[1], and Ekaterina Pomjakushina[1,†]

[1] Laboratory for Multiscale Materials Experiments, Paul Scherrer Institut, CH-5232 Villigen PSI, Switzerland

[2] Laboratory for Neutron Scattering and Imaging, Paul Scherrer Institut, CH-5232, Villigen PSI, Switzerland

[3] Laboratory for Neutron and Muon Instrumentation, Paul Scherrer Institut, CH-5232 Villigen PSI, Switzerland

[4] Key Laboratory of Polar Materials and Devices (MOE), School of Physics and Electronic Science, East China Normal University, Shanghai 200241, China

[*]soohyeon.shin@psi.ch, [†] ekaterina.pomjakushina@psi.ch

# Figures and captions

**Figure S1**

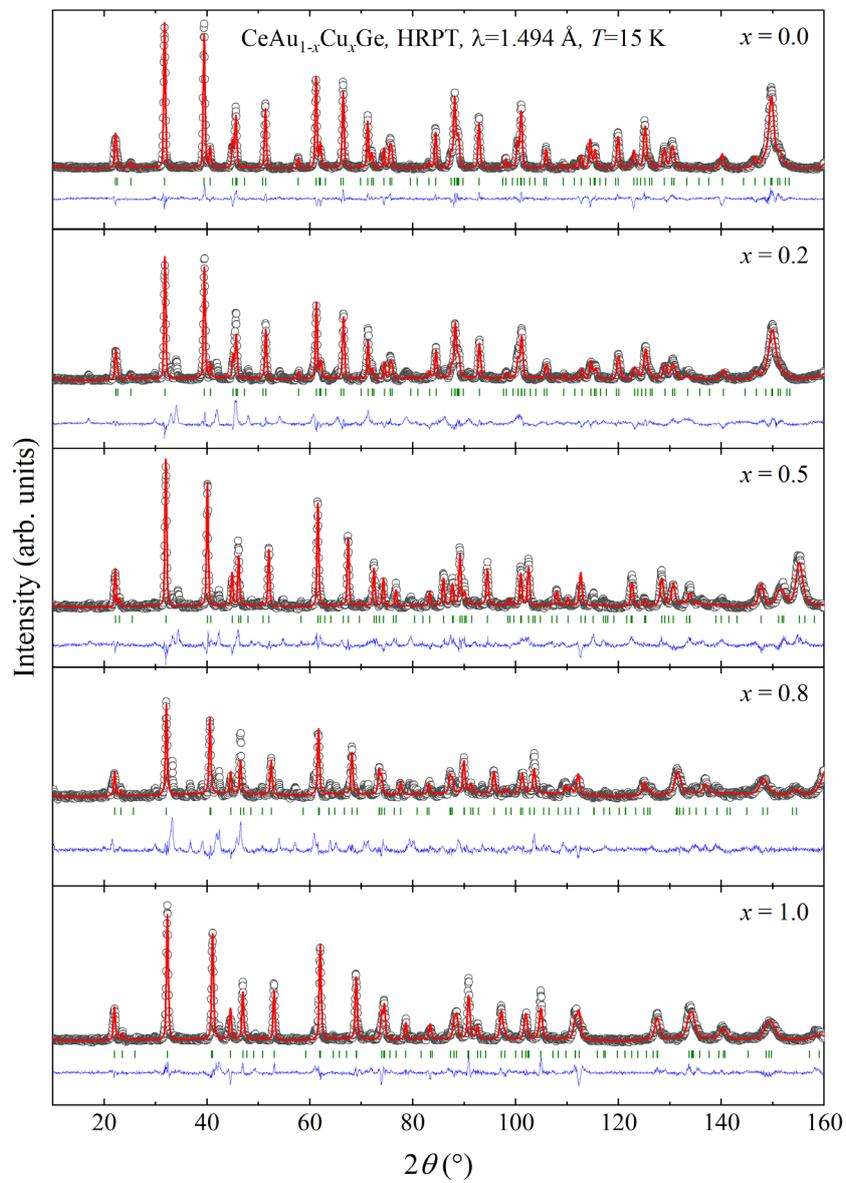

Fig. S1. Structural Rietveld refinement of neutron diffraction patterns of CeAu$_{1-x}$Cu$_x$Ge measured at 15 K.

**Figure S2**

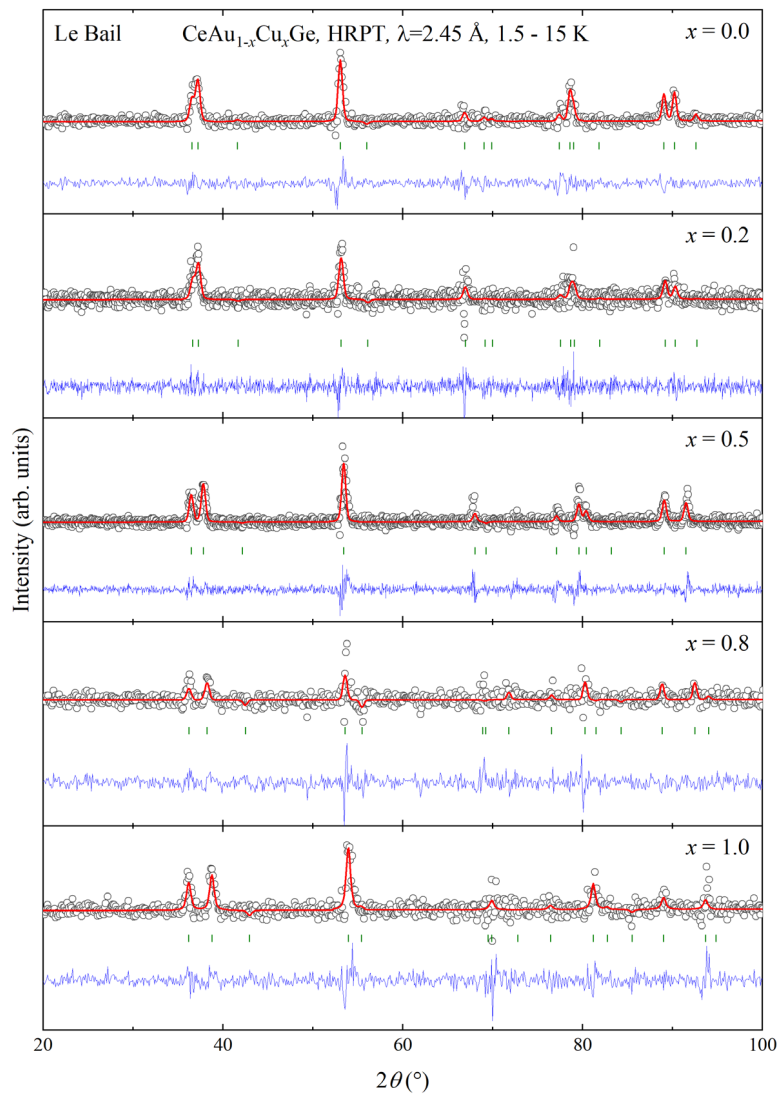

Fig. S2. Le Bail fits of the difference patterns of CeAu$_{1-x}$Cu$_x$Ge with propagation vector $\boldsymbol{k} = 0$. The space groups are indicated in Table S1.

**Table S1**. **Goodness of fit for the fits shown in Fig. S2.**

| $x$ | Space group | $R_{wp}$ | $R_{exp}$ | $\chi^2$ |
|---|---|---|---|---|
| 0.0 | $P6_3mc$ | 84.0 | 84.2 | 1.00 |
| 0.2 | $P6_3mc$ | 84.5 | 84.5 | 1.00 |
| 0.5 | $P6_3/mmc$ | 68.7 | 66.6 | 1.06 |
| 0.8 | $P6_3/mmc$ | 89.7 | 85.1 | 1.04 |
| 1.0 | $P6_3/mmc$ | 84.2 | 83.6 | 1.01 |

**Figure S3**

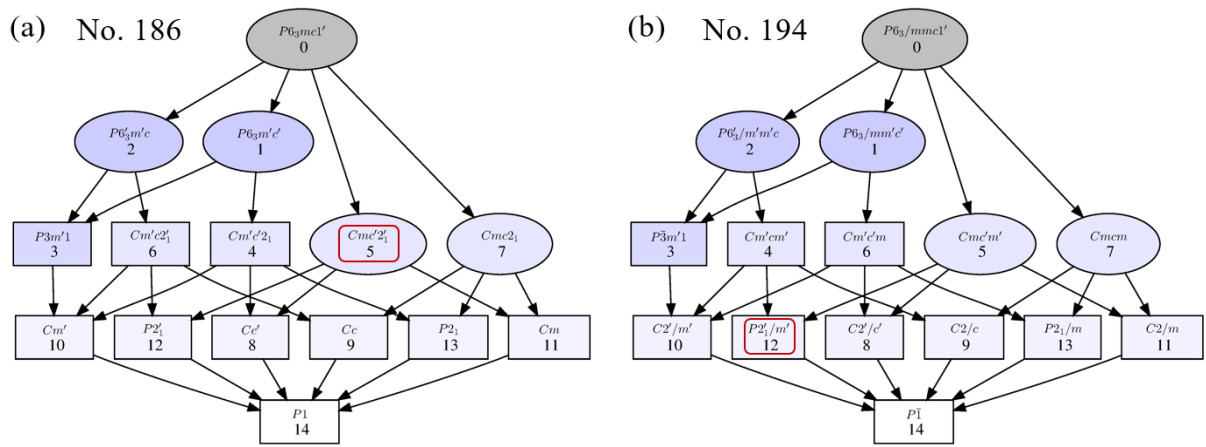

Fig. S3. Graph of subgroups which allow non-zero magnetic moments for propagation vector $k = 0$ in a space group (a) $P6_3mc$ (no. 186) and (b) $P6_3/mmc$ (no. 194). Red box indicates the subgroup giving the best fit of the magnetic structure.

**Figure S4**

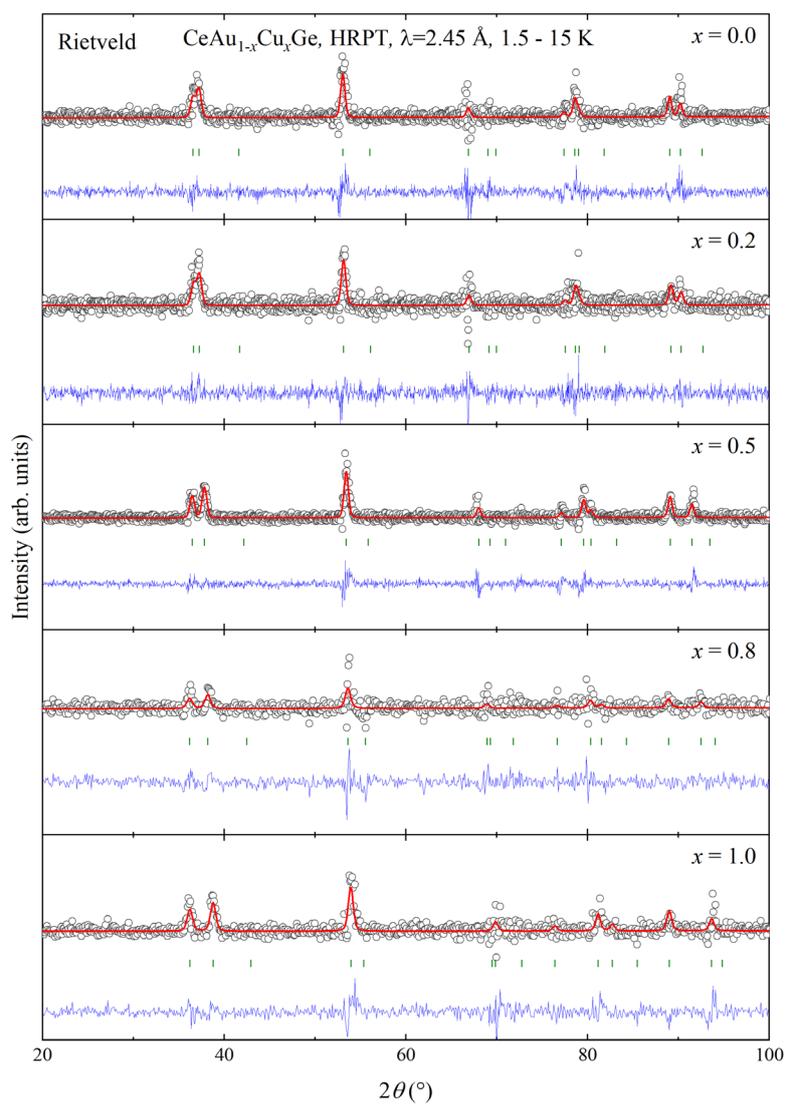

Fig. S4. Rietveld refinement of the difference patterns of CeAu$_{1-x}$Cu$_x$Ge using magnetic models described in Table S3.

**Table S2**. Ordered magnetic moments and goodness of fit for the fits of the difference patterns (1.5 K – 15 K) shown in Fig. S4.

| $x$ | $\mu_{ord}$ ($\mu_B$) | $R_{wp}$ | $R_{exp}$ | $\chi^2$ |
|---|---|---|---|---|
| 0.0 | 1.05(2) | 85.6 | 85.9 | 1.00 |
| 0.2 | 0.95(2) | 84.7 | 84.3 | 1.01 |
| 0.5 | 1.16(1) | 68.4 | 66.4 | 1.06 |
| 0.8 | 1.08(2) | 86.3 | 84.4 | 1.04 |
| 1.0 | 1.11(3) | 84.5 | 83.1 | 1.03 |

**Table S3**. Structure parameters obtained from the fits of neutron diffraction patterns. The unit cell metrics and atomic coordinates are shown for the magnetic space groups. They were fixed by the values refined from the paramagnetic patterns at 15K.

| $x$ | MSG | M W, (x, y, z) | Ge W, (x, y, z) | $a$ (Å) | $b$ (Å) | $c$ (Å) |
|---|---|---|---|---|---|---|
| 0.0 | $Cmc'2_1'$ | 8b, (0, 1/3, 0.8) | 8b, (0, 1/3, 0.2) | 4.4578 | 7.7211 | 7.8540 |
| 0.2 | $Cmc'2_1'$ | 8b, (0, 1/3, 0.8) | 8b, (0, 1/3, 0.8) | 4.4562 | 7.7184 | 7.8464 |
| 0.5 | $P2_1'/m'$ | 2e, (2/3, 1/4, 1/3) | 2e, (1/3, 1/4, 2/3) | 4.3936 | 7.8815 | 4.3936 |
| 0.8 | $P2_1'/m'$ | 2e, (2/3, 1/4, 1/3) | 2e, (1/3, 1/4, 2/3) | 4.3344 | 7.9048 | 4.3343 |
| 1.0 | $P2_1'/m'$ | 2e, (2/3, 1/4, 1/3) | 2e, (1/3, 1/4, 2/3) | 4.2996 | 7.9601 | 4.2996 |

Atomic coordinates of Ce are (0, 0, 0) in all cases and W = 4a and 2a for $Cmc'2_1'$ and $P2_1'/m'$, respectively. MSG and W denote magnetic space group and Wyckoff position, respectively.

**Figure S5**

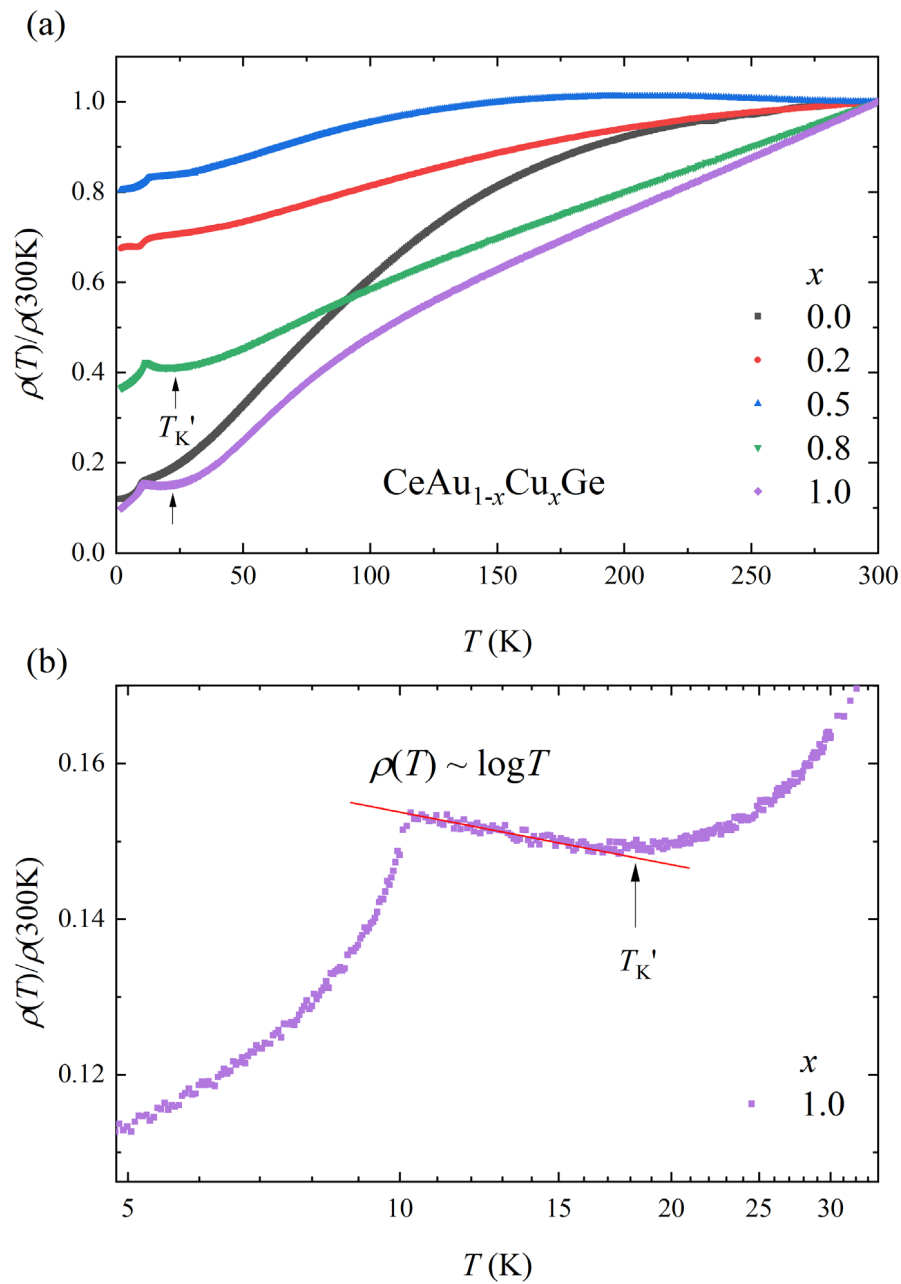

Fig. S5. Electrical resistivity ($\rho(T)$) as a function of temperature of CeAu$_{1-x}$Cu$_x$Ge. (a) $\rho(T)$ normalized by $\rho(300\ \text{K})$, $\rho(T)/\rho(300\ \text{K})$, of CeAu$_{1-x}$Cu$_x$Ge ($x = 0.0, 0.2, 0.5, 0.8, 1.0$) is shown for overall temperature range from 300 to 1.8 K (0.28 K for $x = 0.2$). Black upward arrows indicate the resistivity minimum due to the Kondo effects. (b) Logarithmic temperature dependence of $\rho(T)/\rho(300\ \text{K})$ is plotted for $x = 0.2$ exhibiting a linear behaviour due to the Kondo effect as shown by a red line.

Table S4. **Residual-resistance ratio (RRR), $\rho(300\ K)/\rho(0\ K)$, coefficient of CeAu$_{1-x}$Cu$_x$Ge.**

| $x =$ | 0.0 | 0.2 | 0.5 | 0.8 | 1.0 |
|---|---|---|---|---|---|
| RRR= | 8.3 | 1.6 | 1.2 | 2.7 | 10.6 |